\documentclass[12pt]{iopart}
\usepackage{iopams}
\usepackage{textcomp}
\newtheorem{theorem}{Theorem}

\usepackage{graphicx}
\newcommand{\abs}[1]{\left| #1 \right|}
\newcommand{\bra}[1]{\left\langle #1 \right|}
\newcommand{\braket}[1]{\left\langle #1 \right\rangle}
\newcommand{\ca}[1]{\ensuremath{\mathcal{A}_{#1}}}
\newcommand{\cb}[1]{\ensuremath{\mathcal{B}_{#1}}}

\newcommand{\cd}[1]{\ensuremath{\mathcal{D}_{#1}}}
\newcommand{\cbr}[1]{\left\{ #1 \right\}}
\newcommand{\del}[1]{\left( #1\right)}
\newcommand{\ETh}{\ensuremath{E_{\textrm{Th}}}}		
\newcommand{\fnrm}[1]{\ensuremath{||#1||_\textrm{F}}}	
\newcommand{\floor}[1]{\lfloor #1\rfloor}
\def\hcal{\ensuremath\mathcal{H}}	
\def\hgoe{\ensuremath{H_\textrm{GOE}}}	

\newcommand{\ket}[1]{\left| #1 \right\rangle}
\newcommand{\mean}[1]{\ensuremath{\langle#1\rangle}}
\def\ot{\ensuremath{\mathsf{O}}}				
\def\ob{\ensuremath{\Omega}}				
\def\obo{\ensuremath{\widetilde{\Omega}}}		
\newcommand{\ppt}{\ensuremath{\mathsf{P}}}	
\newcommand{\prob}[1]{\ensuremath{\textrm{P}\left(#1\right)}}	
\newcommand{\pgoe}[1]{\ensuremath{\textrm{P}_\textrm{GOE}\left(#1\right)}}	
\def\phq{\ensuremath{\textrm{Tr}\ppt_k H \qt_k H \ppt_k}}
\newcommand{\pluscross}{\rlap{+}{\texttimes}}
\newcommand{\qt}{\ensuremath{\mathsf{Q}}}	
\def\starn{\ensuremath{\textrm{`}\alpha_1\alpha_2\dots\alpha_{N-1}\textrm{'}}}
\def\shn{\ensuremath{\textrm{S}}}
\def\sij{\ensuremath{\sum_{i\leq k, j>k}}}
\def\Tr{\ensuremath{\textrm{Tr}}}
\newcommand{\tTh}{\ensuremath{t_{\textrm{Th}}}}		
\newcommand{\tR}{\ensuremath{t_{\textrm{R}}}}		
\def\x{0.5}
\begin{document}
	
\title[Chaos in DPE]{Chaos due to symmetry-breaking in Deformed Poisson Ensemble}

\author{Adway Kumar Das \& Anandamohan Ghosh}

\address{Indian Institute of Science Education and Research Kolkata, Mohanpur, 741246 India}
\ead{anandamohan@iiserkol.ac.in}
\vspace{10pt}
\begin{indented}
	\item[]\today
\end{indented}

\begin{abstract}
	The competition between strength and correlation of coupling terms defines numerous phenomenological models exhibiting spectral properties interpolating between those of Poisson (integrable) and Wigner-Dyson (chaotic) ensembles. It is important to understand how the off-diagonal terms of a Hamiltonian evolve as one or more symmetries of an integrable system are explicitly broken.  We introduce a Deformed Poisson Ensemble (DPE) to demonstrate an exact mapping of the coupling terms to the underlying  symmetries of a Hamiltonian. From the maximum entropy principle we predict a chaotic limit which is numerically verified from the spectral properties and the survival probability calculations.
\end{abstract}
\submitto{\JSTAT}
\section{Introduction}
A classical system with $N$ degrees of freedom is deemed integrable if $N$ integrals of motion exist where each of them correspond to an underlying symmetry of the system. An equivalent criteria of integrability is the existence of {\it Lax pairs} $(L,B)$ such that the equation of motion is given by $\frac{d L}{d t} = [B,L]$ \cite{Babelon1}. This isospectral evolution of the Lax operators has an analogous picture in the Heisenberg formulation of quantum mechanics, where time evolution of an observable, $\ot$, is given by $\frac{d\ot}{dt} = \frac{i}{\hbar}[H, \ot]$ while $[H, \ot] = H \ot - \ot H$ and $H$ is a time independent Hamiltonian \cite{Bransden1}. Hence $\ot$ becomes a conserved quantity if it commutes with $H$ i.e., $[H,\ot] = 0$. Therefore a quantum mechanical system is likely to be considered as completely integrable if it is governed by a time independent Hamiltonian with adequate number of commutators. However, it is trivial to construct enough commuting partners from the energy state projectors for any Hamiltonian showing that such notion of integrability in quantum mechanics can be misleading  \cite{Yuzbashyan2, Yuzbashyan3}. 

With the development of Random Matrix Theory (RMT), integrability can also be established as a statistical measure following the Berry-Tabor theorem \cite{Berry3}. Other than the harmonic oscillator and related systems, integrable quantum mechanical systems are described by Hamiltonians possessing independent random energy levels.
These Hamiltonians constitute the Poisson ensemble which  must commute with all the members of a complete set of commuting observables, $\ob$. Consequently there exists a common basis $\Theta$, in which all the Hamiltonians in Poisson ensemble become diagonal with i.i.d. random entries and each entry can be assigned a unique set of good quantum numbers. 
Contrarily the spectra of quantum systems that exhibit chaos in the classical limit show fluctuations similar to that of Wigner-Dyson ensembles, where all the off-diagonal terms are present irrespective of the choice of basis. This correspondence has led to the Bohigas-Giannoni-Schmidt (BGS) conjecture \cite{Bohigas1} laying the foundation of Quantum Chaos \cite{Guhr1}.

In many physical systems, the spectral property of the governing Hamiltonian is intermediate between completely integrable  and chaotic \cite{Torres3, Luitz1, Evangelou1}, a behavior demonstrated in many phenomenological models as well \cite{Khaymovich1, Evers1, Das_arxiv1, Tomasi3}. 
Often in such matrix models, it is possible to control the extent of integrability by varying the coupling strength as in Rosenzweig-Porter ensemble (RPE) \cite{Kravtsov1, Khaymovich2} or the range of coupling as in Power Law Banded Random Matrix (PLBRM) \cite{Evers1} or introducing inhomogeneous coupling as in $\beta$-ensemble \cite{Das_arxiv1}. Alternatively one can introduce correlations among matrix elements so that only certain coupling terms are dominant in an irreducible representation e.g. deformed ensembles \cite{Hussein2, Hussein4}. In all these approaches, the underlying symmetries are broken explicitly, while one can also look at spontaneous symmetry breaking \cite{Pato1}. The main focus of this work is to show that the competition between strengths and positions of the coupling terms explicitly breaks the underlying symmetries, consequently the integrability of a Hamiltonian. We introduce a Deformed Poisson Ensemble (DPE), where an integrable Hamiltonian is perturbed with off-diagonal block matrices and demonstrate an one-to-one correspondence between the matrix elements and symmetries of the Hamiltonian. Such a perturbation has chiral symmetry \cite{Beenakker1} and has been used to break the symmetry in deformed ensemble \cite{Hussein2}. We obtain the distribution of such matrices using the maximum entropy principle \cite{Balian1} and show that entropy is maximized in the completely chaotic/ergodic regime. We numerically verify our claim by systematically studying the spectral properties and dynamics of DPE.
The organization of the paper is as follows: we develop a general framework in section~\ref{sec_struc} relating matrix elements with the underlying symmetry of a Hamiltonian. In section~\ref{sec_mep}, we introduce the Deformed Poisson Ensemble (DPE) and obtain the entropy in matrix space. In section~\ref{sec_prop}, we numerically demonstrate that adding chiral perturbations can drive an integrable system to a chaotic regime.
\renewcommand{\x}{0.32}
\section{Hamiltonian Structures under Symmetry Constraints}\label{sec_struc}
\subsection{Block Structures}
Consider an $N$-dimensional Hilbert space over the real field in which a complete set of commuting observables, $\ob \equiv (\ot_1, \ot_2, \dots, \ot_k, \dots, \ot_n)$ is chosen. Any such $\ot_k$ must be symmetric if it commutes with a real symmetric Hamiltonian (theorem~\ref{thm_1}) and has $m_k$ distinct real eigenvalues, where the $i^{th}$ eigenvalue has the multiplicity $\Delta_{i, k}$ such that $\sum_{i = 1}^{m_k}\Delta_{i, k} = N$.  
Let the common eigenbasis of $\ot_k$'s be $\Theta$, in which their matrix representations are diagonal. Now take a random symmetric Hamiltonian, $H_{k+}$, such that $[H_{k+}, \ot_k] = 0$ for a particular $k$. Though two symmetric matrices commute if and only if they are simultaneously diagonalizable, the presence of degeneracy in $\ot_k$ renders $H_{k+}$ to be block diagonal in the $\Theta$ basis:
\begin{equation}
	\label{eq_Hp}
	H_{k+} =
	\left(\begin{array}{cccc}
		\ca{1, k}& 0& 0& 0\\
		0& \ca{2, k}& 0& 0\\
		0& 0& \ddots& 0\\
		0& 0& 0& \ca{m_k, k}
	\end{array}\right).
\end{equation}
Hence the energy spectrum of $H_{k+}$ consists of $m_k$ independent sub-spectra. Similarly if a random Hamiltonian simultaneously commutes with any $l$ out of $n$ members of $\ob$, its energy spectrum is a superposition of $m'$ pure sequences, where
\begin{equation}
	\label{eq_ok_number}
	m' = 1 - l + \sum_{k = 1}^l m_k, \qquad 2\leq m_k\leq N-n+1,\quad 1\leq n \leq N-1.
\end{equation}
Particularly $m' = N$ for $l=n$, i.e. the Hamiltonian commutes with all the elements of $\ob$ such that all the energy states can be labeled by a unique set of good quantum numbers (eigenvalues of $\ot_k$'s). In this scenario all the energy levels become uncorrelated and the system is integrable by definition. 
We denote the integrable Hamiltonians by $H_+$, which belong to the Poisson ensemble and exhibit complete level clustering following Berry-Tabor theorem \cite{Berry3}.

Let us now find the structure of a Hamiltonian that departs from integrability. 
In the absence of any symmetry, such a Hamiltonian belongs to GOE such that its energy spectrum exhibits complete level repulsion typical of chaotic systems \cite{Bohigas1}. 
We can always decompose such a Hamiltonian as $\hgoe = H_{k+} + H_{k-}$ such that $[H_{k+}, \ot_k] = 0$ and $\{H_{k-}, \ot_k\} = 0$ provided $\ot_k$ is invertible (theorem~\ref{thm_2}). 
Thus the symmetry in the block diagonal $H_{k+}$ is broken by the corresponding $H_{k-}$
possessing only off-diagonal blocks: 
\begin{equation}
	\label{eq_Hm}
	H_{k-}=
	\left(\begin{array}{ccccccc}
		0& & & & & &\cb{1, k}\\
		\cb{1, k}^T&0 & & & & & \cb{2, k}\\
		\vdots& & &\ddots & & & \vdots\\
		\vdots& & & & & 0& \cb{m_k-1, k}\\
		\cb{m_k-1, k}^T& & & & & & 0
	\end{array}\right).
\end{equation}
Therefore the presence (absence) of any symmetry results in creation (destruction) of diagonal blocks in a Hamiltonian. The matrices of the form $H_{k-}$ from \ref{eq_Hm} are crucial to construct the DPE proposed in section~\ref{sec_mep}.

Above ideas can be easily illustrated in the case of two-level systems. Since any non-trivial $2\times 2$ operator has no degeneracy, $\ob$ has only one element, $\ot_1$. If we re-scale and recenter the energy axis, it is convenient to write $\ot_1\equiv \sigma_z$, where $\sigma_{x, y, z}$ are Pauli matrices. 
Any general $2\times 2$ real random Hamiltonian can be uniquely decomposed as $\hcal = uI + v\sigma_z + w\sigma_x$  where $u, v, w$ are random variables. 
The integrable part of the Hamiltonian can be expressed as $H_+ = uI + v\sigma_z$  satisfying $[H_+, \sigma_z] = 0$ while $H_- = w\sigma_x$ such that $\{H_-, \sigma_z\} = 0$. The block structures of two-level system are trivial as shown in the schematics of Figure~\ref{fig_1}. The term $uI$ in $\hcal$ shifts the energy levels of $\hcal$ by $u$, while the spectral correlation of $\hcal$ is controlled by the relative strength of $v$ and $w$. Thus the effective Hamiltonian governing any two-level system can be written as $\hcal = z\sigma_z + \lambda x\sigma_x,$ where $\lambda\in\mathbb{R}$.
Choosing $x, z$ from Normal distribution precisely generates an ensemble of random $\hcal$ belonging to $2\times 2$ RPE \cite{Rosenzweig1}. Thus the notion of symmetry breaking is equivalent to the interplay of disorder in the diagonal and off-diagonal terms inducing a phase transition in the spectral properties \cite{Pandey2, Kravtsov1, Pino1}. The density of NNS for $\hcal$ parameterized by $\lambda$ has been derived in a number of studies \cite{Huu-Tai1, Berry1, Das1}, which exhibits level clustering or repulsion as $\min\{\lambda, 1/\lambda\}\to 0$ or 1 respectively.
\subsection{Orthogonal Operators}\label{sec_orth} 
The number of configurations of the complete set of observables, $\ob$ grows as $2^{N-2}$ due to the constraints in \ref{eq_ok_number}. So it is not possible to examine all of them in order to understand the general structure of a symmetry constrained Hamiltonian. In this work we consider the set of orthogonal operators, $\obo$, with $N-1$ elements ordered as
\begin{equation}
\label{eq_ot_k}
\ot_k(i, j)=\cases{-\delta_{ij},~\quad 1\leq i\leq k\\\delta_{ij},\qquad k<i\leq N.}
\end{equation}

Now any real symmetric Hamiltonian can be partitioned into four non-overlapping blocks as $\left(\begin{array}{cc}
	\ca{k}^{(k\times k)}& \cb{k}^{(k\times k')}\\
	\cb{k}^T& \cd{k}^{(k'\times k')}
\end{array}\right)$ for any $k$ where $k' = N-k$. Then $\ca{k}$ and $\cd{k}$ are non-zero in the block diagonal Hamiltonian, $H_{k+}$ commuting with $\ot_k\in\obo$ while the anticommuting Hamiltonian, $H_{k-}$, will comprise of non-zero $\cb{k}$ and $\cb{k}^T$. 
It should be noted that $|N - 2k|$ number of eigenvalues of such $H_{k-}$ are exactly 0 while the rest comes in $\pm$ pairs, a property known as chirality. Ensemble of $H_{k-}$ with normally distributed entries is known as the Chiral Orthogonal Ensemble of topological order $k$ (ChOE(k)) \cite{Beenakker1, Mondal2}. In deformed ensemble, $H_{k-}$ from ChOE(k) is added to a direct sum of two GOE blocks to tune the presence of a symmetry. In section~\ref{sec_mep}, we add the same perturbation to the Poisson ensemble to produce the DPE.
\subsection{Correspondence between Symmetry and Coupling}
\begin{figure}[t]
	\centering
	\includegraphics[width=0.75\textwidth, trim = {0 307 200 290}]{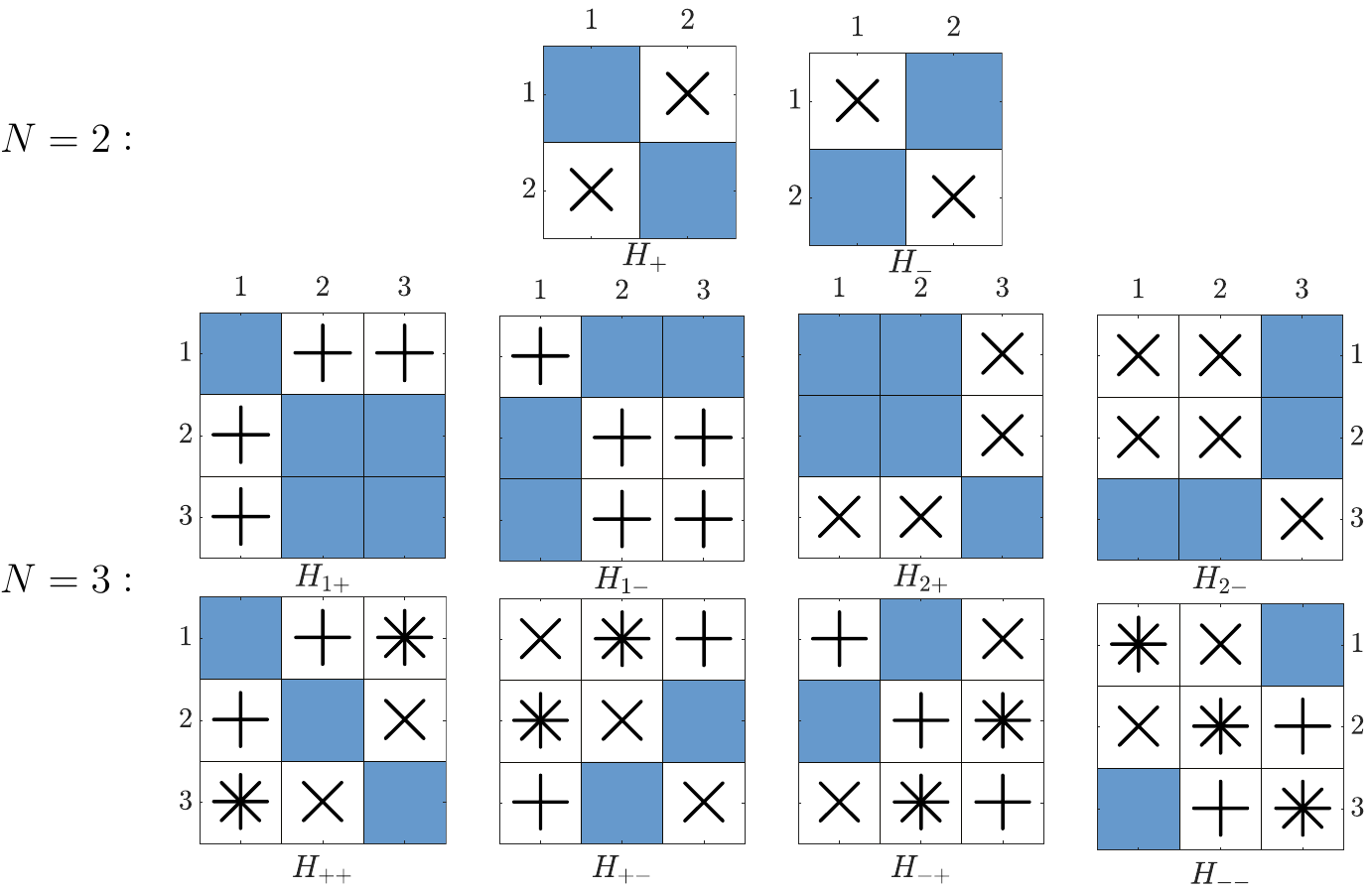}
	\caption{\textbf{Schematic structures of $N\times N$ Hamiltonians due to symmetry constraints}. $\mathbf{N = 2}$: Matrices of the form $H_\pm$ are shown where $[H_+, \ot_1] = 0$, $\{H_-, \ot_1\} = 0$ and $\ot_1\equiv \sigma_z$.
		$\mathbf{N = 3}$: Matrices in the first row are of the form $H_{k\pm}$ ($k = 1, 2$), where $[H_{k+}, \ot_k] = 0,\; \{H_{k-}, \ot_k\} = 0$ and $\ot_k$'s obey \ref{eq_ot_k}. Matrices in the second row are of the form $H_\mathcal{S}$, where $\mathcal{S}$ is a string following \ref{eq_string}. 
		The elements constrained to be 0 are marked $\times$ or $+$, such that any element doubly constrained to be 0 is marked by \pluscross. Only the shaded positions correspond to non-zero elements.
	}
	\label{fig_1}
\end{figure}

For real 3-level systems, we have $\obo \equiv (\ot_1, \ot_2 )$ where the block diagonal matrices $( H_{1+}, H_{2+} )$ 
and the off-diagonal block matrices $( H_{1-}, H_{2-} )$ satisfy the respective commutation and anti-commutation relations. The four possible Hamiltonians of the form $H_{k\pm}$ are schematically shown in Figure~\ref{fig_1}. If any arbitrary Hamiltonian simultaneously commutes with both $\ot_1$ and $\ot_2$, the resultant Hamiltonian $H_{++}$ is integrable and diagonal. $H_{++}$ can also be viewed as a Hamiltonian generated by the elements common in $H_{1+}$ and $H_{2+}$. Similarly, the Hamiltonian that commutes with $\ot_1$ and anti-commutes with $\ot_2$, is symbolically represented as $H_{+-}$, consisting of elements common in $H_{1+}$ and $H_{2-}$. The four possible  structures of $3 \times 3$ Hamiltonians that satisfy all four combinations of chiral symmetries are shown in Figure~\ref{fig_1}.
Similarly a $N\times N$ random real symmetric Hamiltonian $H_\mathcal{S}$ satisfying some or all possible chiral-symmetries can be assigned the string $\mathcal{S}\equiv \starn$ such that
\begin{equation}
\label{eq_main}
\alpha_k = \cases{\textrm{`+'}\quad \textrm{if}\quad [H_{\mathcal{S}}, \ot_k] = 0 ~\Leftrightarrow \cb{k} = 0\\
\textrm{`$-$'}\quad \textrm{if}\quad \{H_{\mathcal{S}}, \ot_k\} = 0 \Leftrightarrow \ca{k}=0,\; \cd{k}=0}\quad \textrm{where }\ot_k\in \obo.
\end{equation}
If a Hamiltonian $H$ neither commutes nor anticommutes with $\ot_k$, we can write $H = H_1 + H_2$, where $[H_1, \ot_k] = 0$, $\{H_2, \ot_k\} = 0$ and follow the above convention to tag each of $H_1$ and $H_2$ separately.


In general, it is possible to uniquely determine which set of symmetries are conserved or violated for each and every element of a Hamiltonian, $H$. Particularly if $H_{ij}$ and $H_{ji}$ ($i \leq j$) are the only non-zero elements, then $H$ can be uniquely mapped to the symbolic string
\begin{equation}
	\label{eq_string}
	\mathcal{S} = \textrm{`}\underbrace{++\dots+}_{1 \textrm{ to } i-1} \underbrace{--\dots-}_{i \textrm{ to } j-1} \underbrace{++\dots+}_{j \textrm{ to } N-1}\textrm{'},\quad i\leq j.
\end{equation}
The symmetry properties of a Hamiltonian constrains the number of allowed strings significantly less than $2^{N-1}$ even though $\mathcal{S}$ has $N-1$ characters. The presence of any off-diagonal element implies that at least one symmetry is broken i.e. there is at least one `$-$' in $\mathcal{S}$. However any arbitrary configuration, e.g. `$\cdots +-++--+\cdots$' is not allowed as any string not conforming to \ref{eq_string} will result in a zero matrix [a graphical proof is given in theorem~\ref{thm_3}]. This restricts the number of admissible strings of length $N-1$ to $N(N-1)/2$. Not surprisingly this is exactly the number of off-diagonal elements in a $N\times N$ matrix.

The real symmetric matrices with 0 on the diagonal and labeled by an admissible $\mathcal{S}$ form a subset of Gell-Mann bases. 
If a random symmetric Hamiltonian contains multiple off-diagonal elements, we can assign different $\mathcal{S}$ to each element and identify the symmetry properties w.r.t. $\obo$ from the characters common to all such strings. 
For example, take a $7\times 7$ Hamiltonian, $H$, whose non-zero elements in the upper-triangle are $H_{2,6}$ and $H_{4,7}$. 
Using \ref{eq_string}, we can assign the following strings to $H_{2,6}$: $\mathcal{S}_{2,6} = \textrm{`}+----+\textrm{'}$ and to $H_{4,7}$: $\mathcal{S}_{4,7} = \textrm{`}+++---\textrm{'}$. 
Upon identifying the common characters, we can immediately conclude that $H$ always commutes with $\ot_1$ while anticommutes with $\ot_{4}$ and $\ot_{5}$. 
The manifestation of uncommon symmetries, i.e. $\ot_{2}$, $\ot_{2}$ and $\ot_{6}$ depends on the relative strength of $H_{2,6}$ and $H_{4,7}$ (e.g. if $H_{2,6}\gg H_{4,7}$, then $H$ commutes with $\ot_6$ while anticommutes with $\ot_{2}$ and $\ot_{3}$). This illustrates the ability of our construction to identify the symmetry properties of any arbitrary random Hamiltonian. In the next section we propose a matrix model to elaborate on the effect of coupling terms in breaking integrability of a system.
\section{Deformed Poisson Ensemble}\label{sec_mep}
For an ensemble of symmetric matrices, $H$, maximizing the Shannon entropy, $\shn = -\int dH\;\prob{H}\log\prob{H}$, results in GOE under the constraints that $\prob{H}$ is normalized and ensemble averaged norm of $H$ is finite \cite{Balian1}. Similar maximum entropy principle in the presence of chiral perturbation of block diagonal matrices generates the deformed ensemble \cite{Hussein2, Hussein3, Hussein4, Hussein5}. The deformed ensemble and its modifications are used to study symmetry breaking in various physical systems, e.g. isospin mixing \cite{Guhr3}, vibrations of crystal block \cite{Carvalho1}, small-world networks \cite{Carvalho2} etc.
In this work we consider Hamiltonians of the form
\begin{equation}
	\label{eq_H_1}
	H = H_+ + V_k
\end{equation}
where $H_+$ is a diagonal matrix and $V_k$ has non-zero entries only in the off-diagonal blocks, $\cb{k}$ and $\cb{k}^T$. We will now show that maximizing $\shn$ of such Hamiltonians subjected to chiral perturbations generates the Deformed Poisson Ensemble (DPE).

The perturbation in $H$ from \ref{eq_H_1} can be expressed as,
\begin{equation}
	\label{eq_Vk}
	V_k = \ppt_k H\qt_k + \qt_k H\ppt_k
\end{equation}
where $\qt_k = \mathbb{I} - \ppt_k$ and $\ppt_k = \sum_{i = 1}^{k}\ket{i}\bra{i}$ are projectors defined over the basis of unit vectors, $\ket{i}$'s. Let, $\fnrm{A} = \sqrt{\Tr A^2}$ be the Frobenius norm of the matrix $A$. Then squared norm of $V_k$ can be written as,
\begin{equation}
	\label{eq_Vk_norm}
	\fnrm{V_k}^2 = 2\phq = 2\sij H_{ij}^2
\end{equation}
Since $\prob{H}$ should be normalized while norms of $H$ and $V_k$ should have finite ensemble averages, the ensemble of $H$ has the following properties,
\begin{equation}
	\label{eq_constraint}
	\mean{\mathbb{I}} = 1,\quad \mean{\Tr H^2} = \mu,\quad \mean{\phq} = \nu
\end{equation}
where $\mean{A} \equiv \int dH\; \prob{H}A$ is the expectation of $A$ over the ensemble of $H$ 
and $\mu, \nu$ are some constants. Using variational principle, if we try to maximize $\shn$ subjected to the constraints in \ref{eq_constraint}, we obtain the density of matrices as, 
\begin{equation}
	\label{eq_P_H_0}
	\prob{H} = \frac{1}{Z}\exp(-\alpha\Tr H^2 - \beta\phq),\quad Z = e^{1+\zeta}
\end{equation}
where $Z$ is the normalization constant and $\zeta, \alpha, \beta$ are Lagrange multipliers. By evaluating the integrals in \ref{eq_constraint}, we obtain the following relations,
\begin{equation}
	\label{eq_relation}
	Z = \sqrt{\frac{\pi^{N + k(N-k)}}{\alpha^N(2\alpha + \beta)^{k(N-k)}}},\: \nu = \frac{k(N-k)}{2(2\alpha + \beta)},\: \mu = \frac{N}{2\alpha} + 2\nu
\end{equation}
The number of free parameters in \ref{eq_P_H_0} can be reduced by proper rescaling of the energy axis and without loss of generality, we assume $\alpha = 1/2$. Then the distribution of matrices can be conveniently expressed as,
\begin{equation}
	\label{eq_P_H}
	\eqalign{\prob{H(k, \lambda)} = \frac{1}{Z}\exp\left( -\frac{1}{2}\sum_{i = 1}^{N}H_{ii}^2 - \frac{1}{2\lambda^2}\sum_{i\leq k, j>k}H_{ij}^2 \right),\quad \lambda = \frac{1}{\sqrt{2(1 + \beta)}}\cr
		Z = (2\pi)^{\frac{N+k(N-k)}{2}}\lambda^{k(N-k)},\: \nu = k(N-k)\lambda^2,\; \mu = N + 2\nu}
\end{equation}
with the Shannon entropy
\begin{equation}
	\label{eq_H_shn}
	\eqalign{\shn(k, \lambda) &= -\int dH(k, \lambda) \prob{H(k, \lambda)} \log\left(\prob{H(k, \lambda)}\right)\\
		&= (1+\log 2\pi)\frac{N + k(N-k)}{2} + k(N-k)\log\lambda.}
\end{equation}
Thus we have obtained the Hamiltonians, $H(k, \lambda)$ such that $\shn(k, \lambda)$ is maximum subjected to the constraints in \ref{eq_constraint}. Moreover \ref{eq_P_H} implies that such Hamiltonians can be expressed as,
\begin{equation}
	\label{eq_H}
	H(k, \lambda) = H_+ + \lambda H_{k-}
\end{equation}
where $H_+\in$ Poisson ensemble and $H_{k-}\in$ ChOE$(k)$ and the collection of $H(k, \lambda)$ defines the Deformed Poisson Ensemble (DPE). Here $\lambda$ controls the overall perturbation strength such that $H(k, \lambda)$ becomes integrable as $\lambda\to 0$ whereas the chiral part, $H_{k-}$ dominates for $\lambda \gg 1$. 
It is easy to see from \ref{eq_H} that $\shn(k, \lambda)$ is maximum w.r.t. $k$ at $k = N/2$. This suggests that DPE is maximally chaotic at $k = N/2$ for a given $\lambda$.

The structure of $H(k, \lambda) \in$ DPE is very intriguing. Firstly, all the symmetries w.r.t. $\obo$ are broken here in a non-trivial manner for any $k$. This can be easily realized if each off-diagonal element of $H(k, \lambda)$ is assigned a string following \ref{eq_string}.
Secondly, as the off-diagonal block $\cb{k}$ with dimension $k\times (N-k)$ is non-zero in $H(k, \lambda)$, each of the first $k$ diagonal terms is coupled to the last $N-k$ diagonal terms. Similarly each of the last $N-k$ terms is coupled to the first $k$ elements in the diagonal. 
Hence the coordination number of the first $k$ and last $N-k$ terms are $N-k$ and $k$ respectively. 
Thus for $k=N/2$, all the diagonal terms are connected to $N/2 (= \mathcal{O}(N))$ terms and should result in a completely chaotic behavior. 
Contrarily for $k\neq N/2$, spatial inhomogeneity in the coupling can result in a deviation from the chaotic behavior. 
Again due to the symmetry $\ot_{k} = -\ot_{N-k}$ evident from \ref{eq_ot_k}, the spectral properties of DPE are same for $k$ and $N-k$, i.e. we need to investigate DPE for $k = 1$ to $N/2$.
From \ref{eq_P_H}, we observe that the norm of the off-diagonal part, $2\nu \propto k(N-k)$ scales with $N$ so we denote $k$ as 
\begin{equation}
	\label{eq_k_scale}
	k = \floor{(N/2)^\gamma}
\end{equation}
and observe the spectral properties as a function of $\gamma \in [0,1]$.
In the next section we fix the perturbation strength at $\lambda = 1$ and vary $\gamma$ to show the emergence of chaotic phase.

\section{Spectral Properties}\label{sec_prop}
\begin{figure}[t]
	\centering
	\includegraphics[width=\textwidth, trim = {0 345 0 345}]{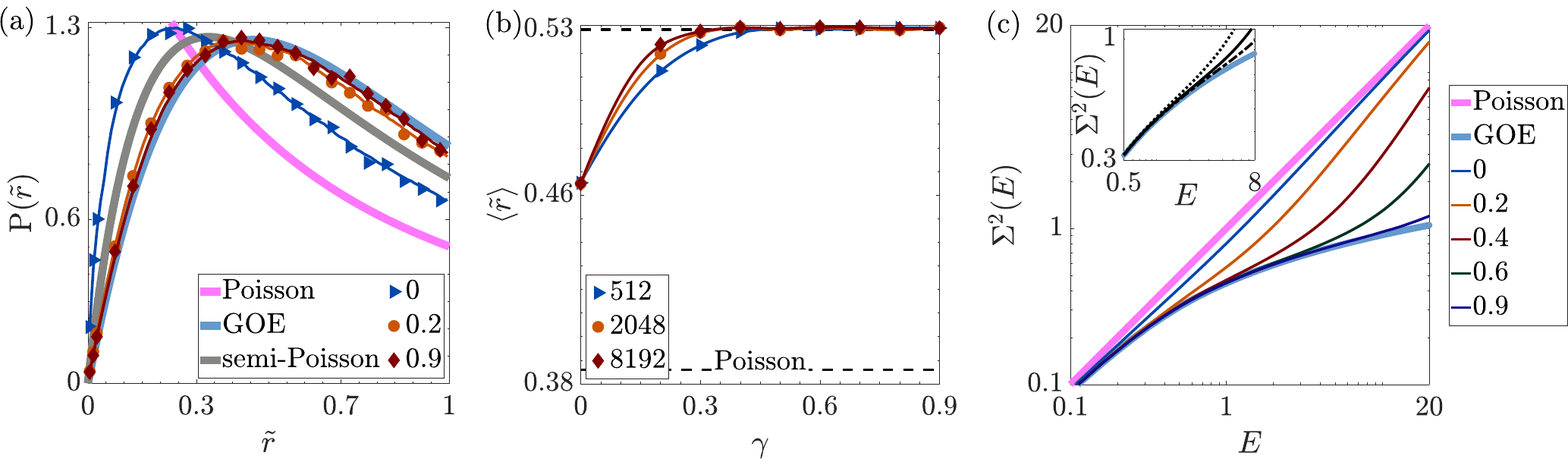}
	\caption{\textbf{Energy level correlation for $H(k, 1)$:} (a) PDF of RNNS for $N = 8192$ averaged over 125 disorder realizations for different $\gamma$, where $k = \floor{(N/2)^\gamma}$. (b) Ensemble average of RNNS for different $N$ as a function of $\gamma$. (c) Number variance, $\Sigma^2(E)$ as a function of energy, $E$ for $N = 8192$ while varying $\gamma$. Inset shows number variance for $\gamma = 0.7$ and $N = 512, 2048, 8192$ denoted by dashed, solid and dot-dashed lines respectively. In (a) and (c), bold curves denote analytical formulae for the limiting cases.
	}
	\label{fig_2}
\end{figure}

\paragraph{\bf Density of States (DOS):}The chiral perturbation present in the Hamiltonians from DPE manifests itself in the shape of the DOS. Recall that $|N-2k|$ eigenvalues of $H_{k-}$ must be 0, which in turn splits the energy spectrum of $H(k, \lambda)$ into three separate bands: a central band with $|N-2k|$ energy levels centered around $E = 0$ and two edge bands of lengths $k$ with positive and negative energies. Since extreme eigenvalues generically possess different correlation than the bulk spectrum, we consider only the central energy band for spectral analysis. The density of central $|N-2k|$ eigenvalues follows Gaussian distribution with a width not changing appreciably with system size, $N$. Consequently for any $k$, mean level spacing of DPE is $\sim 1/N$, a behavior observed in many physical systems governed by sparse Hamiltonians \cite{Brody2, Schiulaz1}. 
\paragraph{\bf Ratio of Nearest Neighbor Spacing (RNNS):} The $i^{th}$ RNNS is defined as $\tilde{r_i} = \min\cbr{r_i, 1/r_i}$ where $r_i = (E_{i+1}-E_i)/(E_i-E_{i-1})$ and $\cbr{E_{i-1}, E_i, E_{i+1}}$ are three consecutive ordered eigenvalues \cite{Oganesyan1}. 
The RNNS statistics captures the short-range correlations present in the energy spectrum of a system  whereas the ensemble averaged value of RNNS, $\mean{\tilde{r}}$, quantifies the degree of repulsion present in a local energy scale \cite{Atas1} . 
In Fig.~\ref{fig_2}(a), we show the PDF of RNNS, $\prob{\tilde{r}}$, for $N = 8192$, averaged over 125 disorder realizations  and choosing middle 60\% of the central energy band. At $\gamma = 0$ (i.e. $k = 1$), $\prob{\tilde{r}}$ is distinct from Poisson or semi-Poisson distributions \cite{Bogomolny5} and approaches GOE as $\gamma \to 1$. In Fig.~\ref{fig_2}(b), we show $\mean{\tilde{r}}$ as a function of $\gamma$ for different system sizes. We observe that $\mean{\tilde{r}}\approx 0.46$ at $\gamma = 0$ for all $N$ (note that $\mean{\tilde{r}} = 0.5$ and $\approx 0.39$ for semi-Poisson and Poisson ensemble respectively).
As we increase $\gamma$, $\mean{\tilde{r}}$ quickly converges to $0.53$ which is the characteristic value for GOE. Thus complete level repulsion as a signature of chaos is evident for $\gamma \to 1$ (i.e. $k\to N/2$).
\paragraph{\bf Number Variance:}The long-range correlations present in the energy spectrum is quantified by the number variance defined as \cite{Backer1},
\begin{equation}
	\label{eq_nvar}
	\Sigma^2(L) = \mean{[\mathcal{N}(L, E) - L]^2}_{\hat{E}, \delta},\quad \mathcal{N}(L, E) = I\del{E + \frac{L}{2}} - I\del{E - \frac{L}{2}}
\end{equation} 
where $\mean{\dots}_{\hat{E}, \delta}$ denotes a local average over an energy window of span $\delta$ centered at $\hat{E}$, $I(E)$ is the cumulative density of the unfolded energy levels and $\mathcal{N}(L, E)$ is the number of energy levels in the window of span $L$ centered at $E$. For GOE, the energy spectrum is rigid so that any two energy levels are always correlated and the number variance exhibits a logarithmic behavior. 
Contrarily in Poisson ensemble, the number of unfolded energy levels are free to fluctuate around their mean position, $\mean{\mathcal{N}(L, E)}_{\hat{E}, \delta} = L$ and the number variance exhibits a linear behavior. 
In Fig.~\ref{fig_2}(c), we show the number variance for $N = 8192$ while varying $\gamma$. We observe that $\Sigma^2(E)$ is almost linear for $\gamma = 0$ indicating that very small correlation is present over long range.
When $\gamma>0$, number variance shows a logarithmic behavior similar to GOE for small energy gaps but starts deviating as we probe longer intervals. This deviation occurs at a larger energy gap as we increase $\gamma$ keeping fixed $N$ as shown in Fig.~\ref{fig_2}(c). Similar effect is observed if we increase $N$ with fixed $\gamma$, as depicted in the inset of Fig.~\ref{fig_2}(c). Such a behavior indicates the presence of an energy scale known as the Thouless energy, $\ETh$. Within an energy interval smaller than $\ETh$, any two energy levels are correlated similar to GOE whereas the correlation is absent for two energy levels separated by a distance larger than $\ETh$, a feature reminiscent of the transport properties in non-interacting systems \cite{Schiulaz1}.

\begin{figure}[t]
	\centering
	\includegraphics[width=\textwidth, trim = {0 345 0 345}]{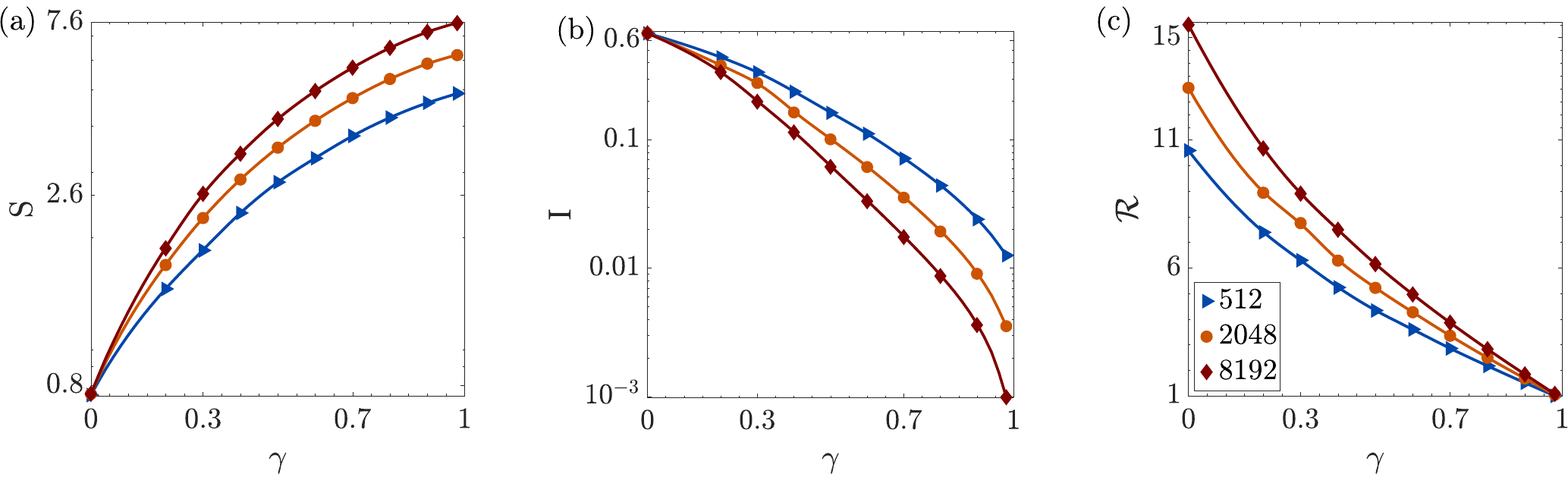}
	\caption{\textbf{Energy state statistics for $H(k, 1)$:} (a) Shannon entropy (b) IPR and (c) relative R\'enyi entropy as a function of $\gamma$ for different system sizes where $k = \floor{(N/2)^\gamma}$. 
	}
	\label{fig_3}
\end{figure}
\paragraph{\bf Localization properties:}
With sufficient evidence that the eigenvalues of DPE indicate the emergence of chaos as $\gamma\to 1$, it is imperative to study the localization properties of the eigenfunctions.
Two commonly studied quantities in this regard are the Shannon entropy, defined as $\shn = -\sum_{i = 1}^{N}P_i\log(P_i)$ and the Inverse Participation Ratio (IPR) defined as I = $\sum_{i = 1}^N P_i^2$ where $P_i = |\Psi_i|^2$ and $\Psi_i$ is the $i^{th}$ component of the eigenstate $\ket{\Psi}$. In Fig.~\ref{fig_3}(a) and (b), we show the Shannon entropy and IPR averaged over middle 60\% of the central band as a function of $\gamma$ for different system sizes, $N$. We observe that for $\gamma = 1$, $\shn\sim \log(0.48 N)$ and I $\sim 3/N$, i.e. the energy states are completely extended in the Hilbert space similar to GOE. On the other hand for any $0<\gamma<1$, we observe that $\shn\sim \gamma\log N$ and I $\sim N^{-\gamma}$, i.e. in the thermodynamic limit ($N\to\infty$), the energy states are extended over an infinite number sites (as $N^\gamma\to\infty$ for $\gamma>0$) which however accounts for a zero fraction of the total Hilbert space volume (as $N^\gamma/N\to 0$ for $\gamma<1$). Contrarily for $\gamma = 0$, both Shannon entropy ($\shn\approx 0.89$) and IPR (I $\approx 0.63$) are constants independent of the system size. Equivalently the entropic localization length, $d_N\equiv 2.07e^\shn$, is $\mathcal{O}(1)$ and $\mathcal{O}(N)$ for $\gamma = 0$ and $\gamma = 1$ respectively, clearly indicating a transition from localized to extended states.
\paragraph{\bf Ergodic properties:}
Now we need to understand whether the extended states for $\gamma>0$ are ergodic or not.
Let $\ket{\Psi_i^j}$ be the $i^{th}$ eigenvector of the $j^{th}$ disordered realization of DPE. Then the relative R\'enyi entropy between two eigenstates with similar energy densities but from different realization of the Hamiltonians is defined as
\begin{equation}
	\label{eq_rr}
	\mathcal{R} = -2\log\del{\sum_{k = 1}^N\abs{\Psi_i^j(k)\Psi_{i+1}^{j^\prime}(k)}}
\end{equation}
which measures the similarity between the support sets of the eigenstates $\ket{\Psi_i^j}$ and $\ket{\Psi_{i+1}^{j^\prime}}$. In particular, $\mathcal{R}\approx 1$ if both the energy states are completely extended and ergodic while $\mathcal{R}\gg 1$ for non-ergodic states \cite{Das_arxiv1}. In Fig.~\ref{fig_3}(c), we observe that $\mathcal{R}\gg 1$ and increases with $N$ for $\gamma<1$ whereas $\mathcal{R}\sim \mathcal{O}(1)$ for $\gamma\to 1$. Thus DPE can enter the ergodic regime only for $\gamma = 1$.
Now we look into the dynamical aspects of the DPE while varying $\gamma$.
\paragraph{\bf Survival probability:}
Given a Hamiltonian $H$, we choose a unit vector $\ket{j}$ having energy close to the center of the energy spectrum of $H$. We let $\ket{j}$ to evolve under $H$ and monitor the time evolution of its survival probability defined as \cite{Schiulaz1},
\begin{equation}
	\label{eq_surv}
	R(t) = \abs{\braket{j|j(t)}}^2 = \abs{\sum_{k=1}^N \abs{\phi_k^{(j)}}^2 e^{-iE_kt}}^2,\quad \phi_k^{(j)} = \braket{\phi_k|j}
\end{equation}
where $(E_k, \ket{\phi_k})$ is the $k^{th}$ eigenpair of $H$. The qualitative behaviors of the survival probability for DPE is described below in the different regimes.
\begin{figure}[t]
	\centering
	\includegraphics[width=\textwidth, trim = {0 345 0 345}]{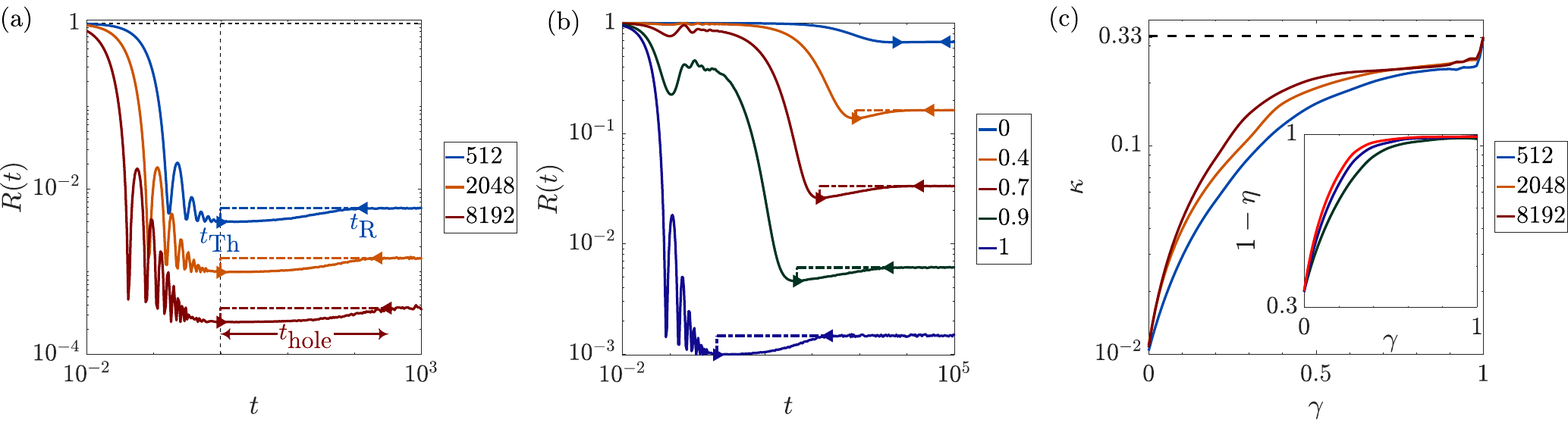}
	\caption{\textbf{Survival probability for $H(k, 1)$:} (a) for different system sizes, $N$ and fixed $\gamma = 1$ where $k = \floor{(N/2)^\gamma}$ (b) for different $\gamma$ and fixed $N = 2048$. In (a) and (b), we mark the Thouless ($\tTh$) and relaxation time ($\tR$) via markers, while the correlation hole ($t_\textrm{hole}$) is shown via dot-dashed lines for each curve.
		(c) $\kappa$ as a function of $\gamma$ for different $N$, where $\kappa = 1 - R(\tTh)/\bar{R}$ measures the relative depth of correlation hole. Inset shows $1-\eta$ vs. $\gamma$ where $\eta$ is the chaoticity indicator.
	}
	\label{fig_4}
\end{figure}

For $\gamma = 1$, the time evolution of survival probability for various $N$ is shown in Fig.~\ref{fig_4}(a). We observe that $R(t)$ decays with time and exhibits a rapid oscillation until reaching the minima at $t = \tTh$, known as the Thouless time \cite{Schiulaz1}. Moreover $\tTh\sim \mathcal{O}(1)$ is system size independent indicating that for $\gamma = 1$, Hamiltonians from DPE are so dense that an initially localized state can quickly spread over the entire Hilbert space irrespective of its dimension. For $t>\tTh$, survival probability keeps increasing until saturating at $t = \tR$, known as the relaxation time, which roughly scales as $\sqrt{N}$. So we observe a dip below the asymptotic value of the survival probability, $\bar{R}$, for a finite time interval, known as the correlation hole, $t_\textrm{hole}\equiv \tR - \tTh$. The existence of a finite correlation hole is a direct manifestation of the spectral rigidity \cite{Torres4}.
For $\gamma<1$, the oscillations present in $R(t)$ for $t<\tTh$ is damped with decrease in $\gamma$ values. We observe the correlation hole at all values of $\gamma$ but its relative depth w.r.t. $\bar{R}$ decreases with $\gamma$. To quantify this effect, we calculate the relative depth of the correlation hole $\kappa = 1 - R(\tTh)/\bar{R}$ \cite{Torres4}. From Fig.~\ref{fig_4}(c), we observe that $\kappa$ attains the maximum at $\gamma = 1$, where $\kappa \approx 1/3$ similar to GOE, reflecting the fact that DPE is completely ergodic and chaotic. For $0<\gamma<1$, $\kappa$ decreases with $\gamma$ and increases with $N$ while $\kappa$ becomes minimum and $N$-independent at $\gamma = 0$. These behaviors are consistent with those of number variance reflecting the fact that long-range correlations among energy levels dies out as $\gamma$ is lowered. Similarly the nature of correlation present in a local energy window is already analyzed via $\mean{\tilde{r}}$. An alternative method to quantify short-range correlation is the chaoticity indicator defined as \cite{Jacquod1, Torres4},
\begin{equation}
	\label{eq_eta}
	\eta = \frac{\int_{0}^{s_0}ds\del{\prob{s} - \pgoe{s}}}{\int_{0}^{s_0}ds\del{\textrm{P}_\textrm{Poisson}(s) - \pgoe{s}}} \approx \frac{\int_{0}^{s_0}ds \prob{s} - 0.16109}{0.215726}
\end{equation}
where $s_0\approx 0.472913$ is the $1^{st}$ intersection point of the PDF of NNS of Poisson and GOE. In the inset of Fig.~\ref{fig_4}(c), we observe that $1-\eta\to 1$ for $\gamma\to 1$ as expected in a chaotic regime. Thus $\kappa$ extracted from the survival probability together with $\eta$ bears the dynamical signature of departure from integrability in case of DPE.
\section{Conclusions}\label{sec_concl}

In this work we observe the emergence of chaotic behavior when an integrable Hamiltonian $H_+$ is perturbed with a chiral block $H_{k-}$ defining a new matrix ensemble called DPE. The departure from integrability can be understood from the one-to-one correspondence between the symmetries of a Hamiltonian and its coupling terms.
For $k = N/2$, each diagonal term in DPE is coupled to $\mathcal{O}(N)$ other diagonal elements such that all possible symmetries are equally broken and results in a chaotic regime. We also prove this by showing that Shannon entropy in the matrix space is maximized exactly at $k = N/2$. Contrarily for $k < N/2$, there are spatial inhomogeneities in the coordination numbers of the onsite terms such that the chaotic behavior gets suppressed. We confirm that the energy states are completely extended and ergodic at $k = N/2$ by systematically studying the spectral properties. By expressing our tuning parameter as $k = \floor{(N/2)^\gamma}$, we identify the following regimes in DPE:
\begin{itemize}
	\item $\gamma = 0$: intermediate short-range correlation different from that in semi-Poisson ensemble while the energy states appear to be completely localized.
	\item $0<\gamma<1$: energy levels separated by distances smaller than $\ETh$ experience complete repulsion while energy states are non-ergodic and extended.
	\item $\gamma = 1$: The energy levels are correlated even over long distances while the energy states are ergodic and extended. The Hamiltonian is dense such that an initially localized state can quickly spread over the entire Hilbert space such that the Thouless time is system size independent.
\end{itemize}

In our proposed matrix model, we demonstrate how the competition between strengths and positions of the coupling terms dictate the symmetries of a system. The consideration of chiral perturbation is restrictive but illustrates the exact mapping of symmetry breaking to matrix structures. Turning on any off-diagonal interaction in a symmetric Hamiltonian, the mechanism of breaking of integrability must be similar but the existence of such an exact mapping is difficult to establish with respect to an arbitrary complete set of commuting observables. 
Any generalization of our framework will be useful with the possibility of identifying the symmetries in a Hamiltonian governing many-body systems undergoing transition from localized to extended phase.

\vskip .1in
{\bf Acknowledgment:} We would like to thank P. K. Mohanty and S. Lal for careful reading of the manuscript and many helpful discussions. AKD is supported by INSPIRE Fellowship, DST, India.

\vskip .1in
{\bf Author contribution statement} AKD and AMG performed the analysis and wrote the paper.
\newpage
\appendix
\setcounter{figure}{0}
\setcounter{table}{0}
\section{Supporting Results}
\begin{theorem}\label{thm_1}
	A real matrix must be symmetric if it commutes with a random real symmetric matrix.
\end{theorem}
{\it Proof:} Given a real matrix $\ot$, let $H$ be a real symmetric random matrix such that $[H, \ot] = 0$. We can decompose $\ot$ into a symmetric and antisymmetric part as $\ot = A+B$, where $A = A^T, B = -B^T$. Then
\begin{equation}
	\eqalign{[H, \ot] &= HA+HB-AH-BH = 0 \cr
		[H, \ot] &= AH-BH-HA-HB = 0}
\end{equation}
Adding and subtracting last two lines, we get $[H, A] = 0$ and $[H, B] = 0$, thus
\begin{equation}
	[H, B]_{ij} = 0\Rightarrow (HB)_{ij} - (BH)_{ij} = (HB)_{ij} + (HB)_{ji} = 0
\end{equation}
as $BH = -(HB)^T$. Consequently $(HB)$ is an antisymmetric matrix, implying
\begin{equation}
	(HB)_{ii} = \sum_{k} H_{ik}B_{ki} = \sum_{k\neq i}H_{ik}B_{ki} = 0 \quad (\because B = -B^T\Rightarrow B_{ii} = 0)
\end{equation}
If $B_{ki}$ is non-zero in general, $H_{ik}$ must be 0 for all $k\neq i$, i.e. H must be diagonal. But our initial construction dictates $H$ to be a random matrix. To resolve this contradiction, $B$ must be 0, i.e. $\ot$ must be symmetric. 
\begin{theorem}\label{thm_2}
	Given two symmetric matrices, $H$ and $O$, we can always obtain $H_\pm$ such that $[H_+, O] = 0$ and $\{H_-, O\} = 0$, provided $O$ is invertible.
\end{theorem}
{\it Proof:} Let us decompose $H$ as
\begin{equation}
	H = H_+ + H_-, \quad H_+ = \frac{H+OHO^{-1}}{2},\; H_- = \frac{H-OHO^{-1}}{2}
\end{equation}
We can easily see that
\begin{equation}
	\label{eq_H_decomp}
	\eqalign{
		&O^{-1}H_+O = O^{-1}\left(\frac{H+OHO^{-1}}{2}\right)O = \frac{O^{-1}HO + H}{2} = \frac{OHO^{-1} + H}{2} = H_+\cr
		&\Bigg(\because H = H^T,\; O = O^T\Rightarrow O^{-1}HO = \left(O^{-1}HO\right)^T = OHO^{-1}\Bigg)\cr
		\Rightarrow & O^{-1}H_+O = H_+ \Rightarrow H_+O = OH_+\Rightarrow [H_+, O] = 0
	}
\end{equation}
similarly one can show that $\{H_-, O\} = 0$. Thus, we obtained the $H_\pm$ with desired properties.
\begin{figure}[h]
	\centering
	\includegraphics[width=\textwidth, trim = {0 350 0 350}]{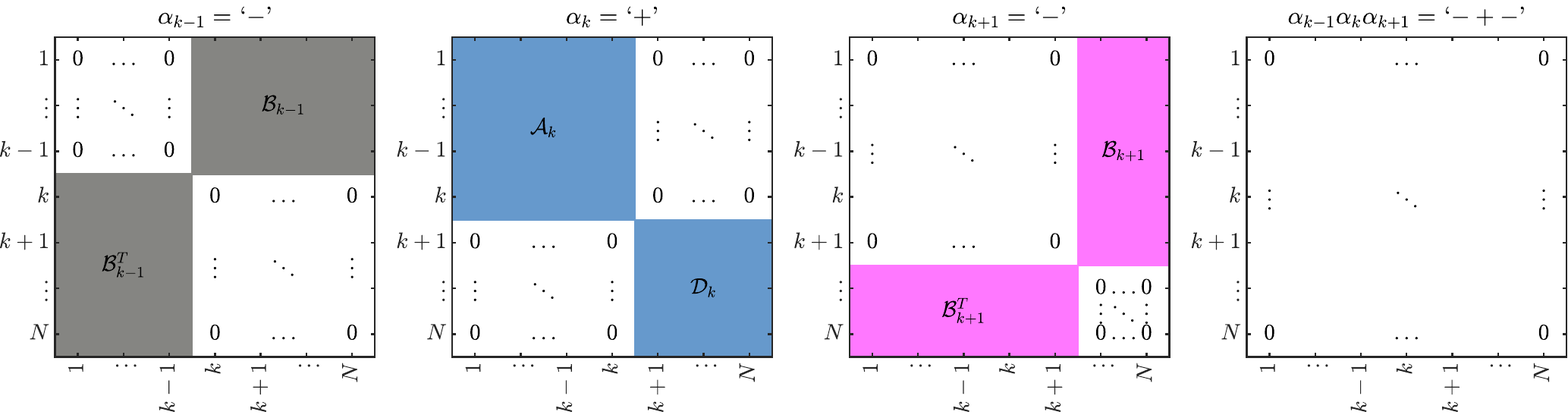}
	\caption{\textbf{Graphical proof} that for a string $\mathcal{S}$ containing a segment `$\cdots-+-\cdots$' must correspond to a zero matrix. 
	}
	\label{fig_5}
\end{figure}
\begin{theorem}\label{thm_3}
	A string $\mathcal{S}$ containing a segment `$\cdots-+-\cdots$' corresponds to a zero matrix.
\end{theorem}
{\it Proof:} Let the positions of $\pm$ in the segment `$\cdots -+-\cdots$' be $k-1, k, k+1$  such that $\alpha_{k-1}=`-'$ , $\alpha_{k}=`+'$ and $\alpha_{k+1}=`-'$ in the string $\mathcal{S}$. Since $\alpha_{k\pm 1}=`-'$, \ref{eq_main} implies that the corresponding matrices have two diagonal blocks containing zeros. Again $\alpha_{k}=`+'$ implies that its corresponding matrix has off-diagonal blocks $\cb{k}=0$ and $\cb{k}^T=0$. These three matrices are schematically shown in Figure~\ref{fig_5}. Now the string $\mathcal{S}=$ `$\cdots-+-\cdots$' necessitates that the structure of $H_{\mathcal{S}}$ should simultaneously conform to all the three structures described above. This is only possible for a zero matrix.

The proof can be easily extended to show that any non-contiguous string of `$-$'  will result in a zero matrix. Hence all admissible strings $\mathcal{S}$ must conform to \ref{eq_string}.

\bibliographystyle{unsrt}
\end{document}